# Rust vs. C for Python Libraries: Evaluating Rust-Compatible Bindings Toolchains

Isabella Basso do Amaral
*University of São Paulo*
isabellabdoamaral@usp.br

Renato Cordeiro Ferreira
*University of São Paulo*
*Jheronimus Academy of Data Science*
renatocf@ime.usp

Alfredo Goldman
*University of São Paulo*
gold@ime.usp

*Abstract*—The Python programming language is best known for its syntax and scientific libraries, but it is also notorious for its slow interpreter. Optimizing critical sections in Python entails special knowledge of the binary interactions between programming languages, and can be cumbersome to interface manually, with implementers often resorting to convoluted third-party libraries. This comparative study evaluates the performance and ease of use of the `PyO3` Python bindings toolchain for Rust against `ctypes` and `cffi`. By using Rust tooling developed for Python, we can achieve state-of-the-art performance with no concern for API compatibility.

## I. Introduction

The Python programming language [1] has been thoroughly used in industry and academia. However, due to its slow interpreter, it serves mostly as glue for lower-level libraries that perform useful algorithms. Although there are methods for accelerating Python programs in Python, its high-level abstractions make it harder to take advantage of machine resources.

Consider the Pythonic `sum` function. A simple benchmark against a poor man's implementation in C (shown in Listing 1) shows a 30x speedup for the latter. The Python `sum` can be easily remedied by using the array programming package `NumPy` [2], which is implemented in C for performance.

Known primarily for its speed, the C programming language [3] has defined the standard binary interface for bindings due to its prevalence, and has been traditionally used for creating universal software packages at the cost of developer sanity, having outstanding undefined behavior issues [4] which prompted generations of additional tooling to remedy many of its design faults.

The C programming language [3], known for its speed, is the standard binary interface for bindings with other programming languages. It has been traditionally used for creating

```
1  static uint64_t
2  sum_list(uint64_t *list, uint64_t n) {
3    uint64_t total = 0;
4    for (uint64_t i = 0; i < n; i++) {
5      total += list[i];
6    }
7    return total;
8  }
```
Listing 1. C implementation of integer sum

TABLE I
Benchmark results for taking the sum of $10^8$ integers.[1]
Execution time is average and standard deviation for 10 executions.

| Implementation | time (ms) |
|---|---|
| Listing 1 (C) | 23.0 ± 0.01 |
| Python `sum` | 618 ± 3.1 |
| `NumPy sum` | 24 ± 0.2 |
| Rust `iter().sum()` | 23.1 ± 0.02 |

universal software packages. However, it has outstanding undefined behavior issues [4]. This prompted generations of additional tooling to remedy many of its design faults.

In the last 20 years, a new generation of systems programming languages emerged from the lessons learned from C, focusing on delivering similar performance. The Rust programming language [5] has become a common alternative to C, and focus on *zero-cost* abstractions. It is built using modern tooling for C, namely the LLVM toolchain [6].

Table I compares Listing 1 with the slow Python `sum`, `NumPy`'s `ndarray.sum()`, and Rust's `iter().sum()` chain (provided by the `std::iter::Sum` trait). The benchmark was run 10 times, taking the sum of 100 million integers. It was executed on Linux 6.14.19 with a Ryzen 7 5800X CPU, using CPython 3.12.8, `NumPy` 2.2.0, and Rust 1.87.0.

This research **evaluates modern tooling alternatives to accelerate Python**. The paper is organized as follows. Section II describes foundational concepts about tooling. Section III delineates the research methodology. Section IV discusses Python FFI mechanisms. Section V compares these approaches. Section VI presents the results of the benchmarks. Section VII analyzes these results. Section VIII proposes answers to the research questions. Section IX addresses limitations of the research methodology. Finally, Section X summarizes the main results and future work.

### A. Research Gap

Table II summarizes papers from the literature that attempt to improve the performance of Python programs using Rust bindings. They were aggregated through Google Scholar search using the keywords "Python", "bindings", "performance", "FFI" and "benchmark".

---
[1]The `NumPy sum` implementation first converts the standard array into a `NumPy` array for fairness.

TABLE II
Research papers reviewed in this study.

| # | Paper | Benchmark Method (execution time) | Source |
|---|---|---|---|
| [9] | D. Teschner *et al.* | Against equivalent tool | GitHub |
| [8] | C. van Amersfoort | Against equivalent tool | GitHub |
| [10] | A. Küsters and W. M. van der Aalst | On domain-specific standard benchmark | GitHub |
| [11] | E. Schubert and L. Lenssen | On MNIST sample | GitHub |
| [7] | E. J. Schofield and D. D. Hodson | Execution time | Unavailable |

Only papers that include benchmark results have been included, which left 5 papers for analysis. Out of those, E. J. Schofield and D. D. Hodson [7] is considered irreproducible as the authors omit all relevant listings and do not provide open source code.

C. van Amersfoort [8] is the only paper that discusses implementation details regarding their use of bindings for Python. The authors abandoned `PyO3` in favor of a domain-specific solution, since the tool lacked automatic interface definitions . However, the authors do not discuss relevant implementation for those bindings.

The remaining papers do not discuss the bindings, even though they used `PyO3`. As such, no reviewed paper demonstrates the reasoning behind those bindings, lacking a complete discussion of bindings strategies.

### B. Reproducibility Concerns

Research software plays a pivotal role in the industry However, it is often developed by researchers that lack real world development experience. As such, the research long-term sustainability is compromised [12]. V. Lenarduzzi, A. Sillitti, and D. Taibi [13] note the near impossible replication of many software studies.

One particular development strategy that appeals to modern scientific standards is that of open source, in which the source code is available to users. Notably, as open-source software is auditable, since it becomes easier to reproduce [14]. This allows collaboration between researchers and developers, which can lead to better software design and performance [15].

All listings and the full experimental setup is open source and can be obtained at https://github.com/isinyaaa/python-ffi

### C. Research question

This research evaluates the feasibility of working consistently in a hybrid-language workflow, enabling high-performance Python through Foreign Function Interfaces (FFI) connecting with systems programming languages. This task has traditionally been done using the C programming language, which can pose great challenges to program correctness.

This paper compares alternative ways to interact with Python, evaluating it with respect to both **performance** and ***developer experience*** (DX). To characterize DX, this research draws inspiration from B. Moseley and P. Marks [16], who differentiates between *essential* and *accidental* complexity.

The implementations are evaluated according to two DX criteria and one performance criterion, which are summarized in the following research questions:

1) **RQ1** [Conciseness]: *What is the accidental complexity of the provided implementation?*
2) **RQ2** [Tooling]: *How hard is the setup of the environment of the provided implementation?*
3) **RQ3** [Performance]: *How fast is the provided implementation against `NumPy`?*

This paper omits memory usage measurements due to limitations in our experimental setup.

## II. Bindings

Each programming language abstracts machine primitives in a different way. Therefore, they expect specific binary layouts for each procedure invocation. A common binary interface is necessary across programming languages, which requires conversions that can also be costly. Such exports are technically known as *Foreign Function Interfaces* (FFI), and commonly known as *bindings*.

### A. C ABI

It is common practice to use the C ABI (Application Binary Interface) as the binary interface between languages. It has two requirements: a register-based calling convention, i.e. using hardware registers as opposed to stack values; and word-sized memory alignment, demonstrated in Fig. 1. Those can be easily achieved in most compiled languages, including Rust. However, linking the exported functions depends on the circumstances, requiring a special treatment for Python.

### B. Compiling and Linking

Labelled memory locations such as functions and global variables are known as *symbols* when treating compiled artifacts. When exported by the compiler, those symbols are available in the *symbol table* for external linkage. They become available in the resulting binary object blob and thus can be reused. It is common to rely on a *linker* to resolve those symbols during compilation.

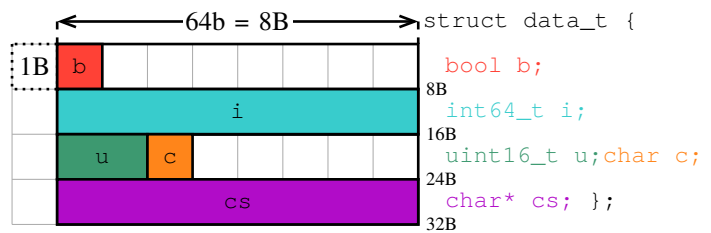

Fig. 1. **C struct padding**[2]. The dotted square denotes the machine address alignment corresponding to 1 byte. The `struct data_t` has five fields with different data types. Even after padding the `bool` with 7 bits, the C compiler will not let `int64_t i` overflow beyond word-alignment. Therefore, 63 bits of padding are used for a single bit. Moreover, a pointer will always occupy an entire word, so there is another padding after our `char c` field.

---
[2]`#include <stdbool.h>` is required to use `bool`, whereas `#include <stdint.h>` is required to use fixed sized integers.

In practice, it is desirable to avoid recompiling the Python interpreter. As such *dynamic loading* is the most common method for bindings, which requires using *shared objects*. Rather than compiling, it is possible to invoke the system calls `dlopen` to open the object and `dlsym` to find a specific symbol. Usually, symbols are known by the implementer, but they can also be dynamically resolved using *libffi*.

### C. Alternatives for Python Bindings

Python has supported dynamic symbol resolution through the ctypes library. However, it may be desirable to manipulate Python objects directly, which is supported through extension modules. They are implemented by compiling a shared library that defines standard symbols recognized by the Python interpreter. They can be parsed upon dynamic loading, exposing Python-native types and functions.

The cffi library allows users to generate Python bindings given the C function signatures automatically. This is the low-level functionality provided by the `PyO3` toolchain [17], which integrates with Rust to provide better DX. The underlying implementation depends on the Maturin build system for Python, which integrates with the Cargo build system for Rust. It can build and link extension modules, replacing the traditional setuptools.

### D. Performance of Python Bindings

There are two methods to expose data through bindings:

1) **M1. *In-situ* conversion.**
   It converts arguments and return values upon function invocation, usually through glue code (not entirely implemented by the user). This is not necessarily a pass-by-value approach, as Python dictionaries and lists still use references and require proper handling.

2) **M2. Specialized constructors.**
   It exposes specialized opaque containers that impose the actual memory boundaries between implementations on the API. Values must be passed by pointer to the opaque struct or class.

Both approaches can be viable depending on the constraints, and are accepted in `NumPy` interfaces. However, specialized constructors are the preferred approach because of the significant difference in binary representation between Python and compiled languages.

## III. METHODOLOGY

To answer the research questions, this paper evaluate the three bindings methods described on Section II.C, namely `ctypes`, `cffi`, and `PyO3`. They will be executed on the same underlying Rust implementation: simple mathematical procedures that are known for poor performance in Python, which are present in `NumPy`.

The DX will be evaluated under conciseness (RQ1), based depend on the expressiveness of the implementation to expose the desired functionality, and tooling (RQ2), based on missing functionality that is required by bindings users.

The performance aspect will be evaluated under performance (RQ3), based on the concerns laid out in Section II.D. In summary, given the heterogeneous Python binary representation, crossing to the C ABI poses several design challenges for any kind of flexibility, often expected in Python-native interfaces such as `NumPy`.

The primary goal of the implementations is to minimize these conversions. As such, there are two implemented versions of each, following the in-situ conversion (M1) and specialized constructor (M2) bindings.

The experiment follows four steps:

1) it compares the two API approaches with the reference, using a standard batched approach to minimize expected overheads for a single procedure call;
2) it breaks up the sample into homogeneous chunks, evaluating function call overhead between the candidates by accumulating batched execution times for the entire sample;
3) it estimates the actual calculation runtime (the lower bound of the graph), which allows estimating the function call overhead in relation to frequency; and
4) it performs a linear regression to find the overhead by call.

## IV. IMPLEMENTATION

To demonstrate Rust's usefulness as a C replacement, it should have reasonable performance for two elementary statistics functions, namely the *arithmetic mean* (`numpy.mean`) and *population standard deviation* (`numpy.std`).

The base implementation for both functions follow the mathematical formulas below:

$$\overline{x}(X) = \sum_{x \in X} \frac{x}{\#X} \tag{1}$$

$$\sigma(X) = \sqrt{\sum_{x \in X} \frac{(x - \overline{x})^2}{\#X}} \tag{2}$$

`NumPy` allows the user to define a *denominator offset* for the variance calculation, so it enables evaluating the sample standard deviation. It also implements its own array types, overloading operators to behave linearly (i.e. operations are performed element-wise), which allows for trivial extension of the arithmetic mean to evaluate the expected value of a distribution. The experiment does not provide flexible interfaces or operator overloads, assuming a minimalist use-case of those functions. Listing 2 shows the functions separated into a `statistics` crate and reused as `stat`.

### A. Traditional Methods for Bindings

By exporting the Rust API, a developer can use the traditional bindings methods (`ctypes` and `cffi`). Since they are implemented in Python, they are the standard option for Python programmers. This technique is useful when the low-level implementation source is not available, but only binaries and the API declaration as a header.

```rust
 1  fn mean(values: &[f64]) -> f64 {
 2      values.into_iter().sum::<f64>()
 3          / (values.len() as f64)
 4  }
 5
 6  fn stddev(values: &[f64]) -> f64 {
 7      let m = mean(&values);
 8      let mut squared_sum = 0.0;
 9      for v in values {
10          let shifted = v - m;
11          squared_sum += shifted * shifted;
12      }
13      (squared_sum
14          / (values.len() as f64)).sqrt()
15  }
```

Listing 2. **Rust implementations for mean and standard deviation.** The functions receive a reference to a slice, which is an abstraction over a pointer to an array that also includes its size. A reference does not need *ownership*, so those functions only *borrow* the array[3].

The C bindings are declared in a separate crate, as it requires different compilation targets. To create a shared library in Rust, a developer must declare the option `lib.crate-type = ["cdylib"]` in `Cargo.toml`.

For each function, the macro `#[unsafe(no_mangle)]` is added to avoid using Rust-specific symbols. Moreover, the statement `extern "C"` in the function signature requests the use the C calling convention, as demonstrated in Listing 3.

To interface with C, the developer has to *cast* memory, i.e., to reinterpret it appropriately into a native Rust type, as demonstrated in the `pointer_to_vec` helper. Since the lifetime of the original memory is not clear, it must copy it in Rust to avoid a double-free. Rust requires using unsafe blocks in order to communicate possible undefined behavior. However such an abstraction does not prevent undefined behavior from occurring, as a caller might provide $n >$ len(values) as an argument, thus provoking a buffer overflow. This is similar to the `sum` function in Listing 1.

It might be preferable to omit the helper entirely, or mark it as `unsafe`. In common practice, it is very undesirable to use unsafe APIs due to the increase in boilerplate and overhead to ensure correctness, which can be reduced by limiting programs to use safe Rust.

### B. Binding with `ctypes`

Listing 4 shows a minimalist module initialization example for the API. The instance must be initialized with the path to the shared library, and it can be scripted.

First, the code converts each `float` instance to a `ctypes.c_double` (aliased as `f64`). It allocates memory using the `ctypes` array constructor syntax.

Second, it takes the product of the type (representing its size) by the array length, getting a new constructor for that specific size. The `c.POINTER` helper function creates a pointer constructor for the specified type, which takes the buffer (array) and produces a Pythonic wrapper for the pointer.

```rust
 1  use statistics as stat;
 2
 3  // <T: Clone> (generic) parameter:
 4  //    defines a parametric constant
 5  //    (type) which specializes the
 6  //    implementation at compile time.
 7  //    `Clone` specifies that the type
 8  //    can be copied recursively.
 9  // unsafe:
10  //    is necessary as the
11  //    `std::slice::from_raw_parts`
12  //    reinterprets memory, which
13  //    might lead to undefined behavior
14  //    if the programmer specification
15  //    is wrong (e.g. actual size != n)
16  fn pointer_to_vec<T: Clone>(
17      values: *mut T, n: u64
18  ) -> Vec<T> {
19      unsafe {
20          std::slice::from_raw_parts(
21              values, n as usize)
22      }.into()
23  }
24
25  #[unsafe(no_mangle)]
26  pub extern "C" fn mean(
27      values: *mut f64, n: u64
28  ) -> f64 {
29      // dereferencing a Vec yields
30      // a slice by default
31      stat::mean(
32          &pointer_to_vec(values, n))
33  }
34
35  #[unsafe(no_mangle)]
36  pub extern "C" fn stddev(
37      values: *mut f64, n: u64
38  ) -> f64 {
39      stat::stddev(
40          &pointer_to_vec(values, n))
41  }
```

Listing 3. **Rust FFI code for interacting with C.** `pointer_to_vec` is defined as a parametric function to enable easy reuse for other sequence types.

Lastly, the code wraps the array pointer with its size into a synthetic `Array` tuple that is only used for typing.

All boilerplate functions can be abstracted to work over polymorphic types.[4]

### C. Classic bindings with `cffi`

To use `cffi` in its intended use case, a developer needs to define a header file for the public interface as shown on Listing 5. This is common practice when providing FFI bindings, since header files are ubiquitous when declaring C APIs and act as documentation.

Fortunately, `cffi` allows using a static library, which we can easily get by adding `staticlib` to the `lib.crate-type` configuration array. This allows simplifying the setup, as the library only needs to be present during the project build step, and not when executing.

---

[3]In Rust's memory model the *owner* will assume responsibility for the value and *drop* it at the end of the scope, freeing its memory. This is not the case when borrowing a value, which only takes a reference.

[4]Check catt.rs library implementation of a converter framework.

```
1  import ctypes as c
2  import typing as t
3  from pathlib import Path
4
5  u64 = c.c_uint64
6  f64 = c.c_double
7  f64_p = c.POINTER(f64)
8  Array = tuple[f64_p, u64]
9
10 lib = c.CDLL(Path(__file__).parent /
   "target/release/libbind_c.so")
11 lib.mean.restype = f64
12 lib.mean.argtypes = [f64_p, u64]
13 lib.stddev.restype = f64
14 lib.stddev.argtypes = [f64_p, u64]
15
16
17 def as_f64(ls: t.Iterable[float])
18      -> t.Iterator[f64]:
19   return (f64(v) for v in ls)
20
21 def array(vs: t.Iterable[f64], n: int)
22      -> Array:
23   return (f64_p((f64 * n)(*vs)),
   u64(n))
```

Listing 4. **Python `ctypes` code for exposing the statistics API.** It uses *iterators* to allow for lazy construction, alleviating the need to create the float values twice.

It is also necessary to define a small script with the C API, which references the header file as well as the compiled binary, as shown on Listing 6.

The cffi setuptools integration works by compiling the source using distutils (i.e. executing the script with FFI.compile()), which is distributed with setuptools. However, it is possible to generate the extension module wrapper code independently of setuptools using FFI.emit_c_code(), as shown on Listing 6. This enables swapping the build system entirely.

Note that in order to use the bindings, it is still necessary to compile the generated extension module. Using setuptools requires packaging the Rust shared object file as well as the header, then specify the desired linking setup used for the extension, shown on Listing 7.

After compiling the extension, the developer can invoke the mean and std functions by importing lib from the compiled _rs_stat module. The extension has built-in type conversion glue code, which is standard for modern FFI generation tools. This adds invisible overhead when calling those functions, as the C type conversions take place and allocate memory for the array *every time each function is called*, as described in the in-situ conversion (M1) binding.

```
1  #include <stdint.h>
2  #include <stdlib.h>
3
4  double mean(
5    double *values, uint64_t n);
6  double stddev(
7    double *values, uint64_t n);
```

Listing 5. **Library header exposing statistics functions.**

```
1  import cffi
2
3  ffibuilder = cffi.FFI()
4  ffibuilder.cdef("""
5  double mean(
6    double *values, uint64_t n);
7  double stddev(
8    double *values, uint64_t n);
9  """)
10 ffibuilder.set_source(
11   "_rs_stat",
12   """
13     #include "header.h"
14   """,
15   # The statement below
16   # is unnecessary for
17   # standalone purposes
18   #libraries=["rs_stat"],
19 )
20
21 if __name__ == "__main__":
22   # setuptools integration
23   #ffibuilder.compile()
24
25   # standalone script
26   ffibuilder.emit_c_code()
```

Listing 6. **Python cffi code for exposing the statistics API.**

A common technique to address this shortcoming is to wrap values in custom types and export constructors, as described in the specialized constructor (M2) binding. This is implemented in Listing 9 and exported as Listing 8. Note that this approach may fail to provide a consistent memory view on the latest values if the type content (either struct or class members) is altered in either implementation after construction.

### D. Maturin Build System

The Maturin CLI tool provides an interactive setup experience, however we still document its behavior for reproducibility purposes.

```
1  [build-system]
2  requires = ["setuptools"]
3  build-backend =
   "setuptools.build_meta"
4
5  [tool.setuptools]
6  ext-modules = [
7    { name = "_rs_stat", sources = [
8      "_cffi.c",
9    ], libraries = [
10     "bind_c",
11   ], library-dirs = [
12     "src/bind_cffi",
13   ]},
14 ]
15 packages.find.where = ["src"]
16 package-data.traditional = [
17   "*.h", "target/release/*.a"
18 ]
```

Listing 7. **`setuptools` build system setup on `pyproject.toml`.**

```
1  struct Array;
2
3  struct Array *
4  array_init(double *, uint64_t);
5  double array_mean(struct Array *);
6  double array_stddev(struct Array *);
```
Listing 8. **Library header exposing a proto-class through an opaque `struct`.** Methods will be prefixed with the record name to follow the C API conventions (unrelated to calling convention).

```
1   #[repr(C)]
2   pub struct Array(Vec<f64>);
3
4   #[unsafe(no_mangle)]
5   pub extern "C"
6   fn array_init(
7     values: *mut f64, n: u64
8   ) -> *const Array {
9     let boxed = Box::new(Array(
10      pointer_to_vec(values, n)));
11    Box::into_raw(boxed)
12  }
13
14  #[unsafe(no_mangle)]
15  pub extern "C"
16  fn array_mean(
17      arr: *mut Array
18  ) -> f64 {
19    mean(&unsafe { &*arr }.0)
20  }
21
22  #[unsafe(no_mangle)]
23  pub extern "C"
24  fn array_stddev(
25      arr: *mut Array
26  ) -> f64 {
27    std(&unsafe { &*arr }.0)
28  }
```
Listing 9. **Rust FFI code exposing a synthetic array.** `Vec` is a dynamically allocated list which is necessary to make our array `Sized` (i.e. fixed size) even though the internal buffer size cannot be determined.

To use Maturin, the developer must define it as our build system in `pyproject.toml`, as demonstrated in Listing 10. They also need a Cargo project, which must also export a shared library. However, Maturin takes care of choosing the correct file depending on the OS, including it with the distribution automatically. It supports many bindings alternatives, including cffi using the `tool.maturin.bindings = "cffi"` option.

### E. PyO3 Extension Module

As previously mentioned, `PyO3` also builds on Maturin, and requires a Python project definition specifying the build system such as in Listing 10. To use `PyO3` in Rust, a developer needs to add it as a dependency with the `extension-module` feature to their Cargo project.

```
1  [build-system]
2  requires = ["maturin==1.8.6"]
3  build-backend = "maturin"
```
Listing 10. **Maturin build system setup on `pyproject.toml`.**

```
1   use statistics as stat;
2   use pyo3::prelude::*;
3
4   #[pyfunction]
5   fn mean(values: Vec<f64>) -> f64 {
6       stat::mean(&values)
7   }
8
9   #[pyfunction]
10  fn stddev(values: Vec<f64>) -> f64 {
11      stat::stddev(&values)
12  }
13
14  #[pymodule]
15  fn bind_pyo3(m: &Bound<'_, PyModule>)
    -> PyResult<()> {
16      m.add_function(
17          wrap_pyfunction!(mean, m)?)?;
18      m.add_function(
19          wrap_pyfunction!(stddev, m)?)
20  }
```
Listing 11. **Rust `PyO3` bindings code.**

Instead of defining the API directly as `extern "C"`, a developer can use `PyO3` helpers to define a Pythonic interface, as demonstrated in Listing 11.

`PyO3` can also wrap Rust `structs` as native Python classes, as shown in Listing 12. This makes the interface more adequate for Python users, and allows easy reuse of the converted types, like in Listing 4.

### V. COMPARISON

This section considers how the **binary conversion** takes place, hinting on their performance characteristics. It discusses the **usability** of the generated APIs and its **tooling**.

Table III summarizes implementation details for each FFI alternative.

```
1   #[pyclass]
2   struct Array(Vec<f64>);
3
4   #[pymethods]
5   impl Array {
6     #[new]
7     fn new(values: Vec<f64>) -> Self {
8       Self(values)
9     }
10    fn mean(&self) -> f64 {
11      stat::mean(&self.0)
12    }
13    fn stddev(&self) -> f64 {
14      stat::stddev(&self.0)
15    }
16  }
17
18  #[pymodule]
19  fn bind_pyo3(m: &Bound<'_, PyModule>)
    -> PyResult<()> {
20    m.add_class::<Array>()
21  }
```
Listing 12. **Rust `PyO3` class definition.**

TABLE III
User features of each FFI implementation discussed.

| FFI Method | Memory Handling | Binding Method |
|---|---|---|
| ctypes | Python | libffi with manual declaration |
| cffi | native | Inferred from the C API |
| PyO3 | native | Inferred from the Rust API |

## A. ctypes

The library exposes C constructor helpers which can be used to specify types. Those can then be used to specify arguments and return types for the API through custom attributes in a c.CDLL instance using libffi, as shown in Listing 4.

1) **Conversion.** The conversions are performed explicitly by the caller in Python, which hints at bad performance.
2) **Usability.** All type conversions create a separate view of the values, and typing errors often lead to unexpected behavior. This makes these declarations very error prone. The conversions can be abstracted in Python, which is significantly more ergonomic than any other declaration method. The one-way nature of the API is taken for granted as a usability concern as a primary limitation of classic FFI methods, i.e. those that do not adopt Python extension modules.
3) **Tooling.** Using cffi only requires a build-system setup for the shared object file, which might be dispensable depending on the use-case (e.g. the library might be installed separately). Interfaces can be typed with no additional cost by bindings creators, however the entire process suffers from lack of automation.

## B. cffi

By using the C header interface for the API (shown in Listing 5), it is possible to auto-generate the Python-C conversions, which are built into the bindings. The bound code is available by importing the generated library (requiring build system integration).

1) **Conversion.** The conversions are performed implicitly per-call. This requires a different export approach in order to enable API specialized constructor (M2) using opaque pointers, shown in Listing 9.
2) **Usability.** By creating the bindings as an extension module, the interpreter is able to catch parameter mismatches. cffi itself guarantees the types are converted sensibly. While exporting C functions is a more general approach to FFI creation, it severely limits the API expressiveness in Python by default. Those bindings can also be supplemented in Python, which is a close approximation to what NumPy offers.
3) **Tooling.** Using Maturin, the C headers can be entirely avoided when working with Rust. However, the cffi method still requires C exports, so all other caveats apply. There is no standard tooling for automatic typing declarations, usually requiring the use of .pyi files with stubs.

## C. PyO3

PyO3 offers Rust macros (code generation) to export functions, enabling Rust-native extension modules while preserving the same working characteristics as cffi with Maturin. The major difference among them being related to **tooling**, where PyO3 closes the gap in creating more ergonomic interfaces for users.

## VI. Benchmarks

The following experiments were executed on Linux 6.14.19 on a Ryzen 7 5800X CPU. They use the Rust 1.87.0 toolchain, CPython 3.12.8 standard single-threaded build and Python dependencies NumPy 2.2.0 and cffi 1.17.1.

Experiments were setup by reading the samples from a file, generating the NumPy result to assert implementation correctness and then timing, each using Listing 13.

The experiments were run 10 times for each sample, using three different random samples. They measure the total time spent on the foreign function calls for each of the 30 individual runs. Benchexec [18] was used for executing the benchmarks in a container with the runexec tool. The experimental setup is available at https://github.com/isinyaaa/python-ffi

The experiments compare the two API approaches, where they either include the binary representation conversion in every call or not, as described in in-situ conversion (M1) and specialized constructor (M2) bindings, respectively.

## A. Serial Executions

The first benchmark compares serial executions. Table IV and Table V show the results along with reference values. As expected, the cffi approach behaves similarly independently of the build system, with differences accounted for uncertainty.

Table IV reveals the expected performance issues with ctypes. Since its a legacy method, the benchmark pursues only the approaches that integrate with Rust and avoid libffi. The subsequent benchmarks compare only NumPy and PyO3, as the other methods have equivalent implementations and results.

```
1  import gc
2  import math
3  import time
4
5  gc.disable()
6  timer = time.perf_counter_ns
7
8  def benchmark(fp, expected, *args,
9                tolerance=0.01):
10     start = timer()
11     actual = fp(*args)
12     end = timer()
13     assert not math.isnan(actual) and
   abs(actual - expected) < tolerance,
14         f"{expected:.3f} != {actual:.3f}"
15     return end - start
```

Listing 13. **Benchmark function using function pointers.** Prior to the benchmark we stop the Python garbage collector to minimize interpreter overheads. The function verifies whether the results is within a 1% tolerance margin from the expected.

TABLE IV
BENCHMARK RESULTS FOR SERIAL RUNS CONVERTING PARAMETERS AT THE CALL SITE, AS DESCRIBED ON IN-SITU CONVERSION (M1).

| Method | mean (ms) | stddev (ms) |
|---|---|---|
| ctypes | 1.978e+05 ± 1.399e+03 | 1.973e+05 ± 1.037e+03 |
| cffi (setuptools) | 7.347e+03 ± 216.20 | 7.717e+03 ± 99.04 |
| cffi (Maturin) | 7.262e+03 ± 60.99 | 7.704e+03 ± 62.76 |
| PyO3 | 6.423e+03 ± 34.70 | 6.965e+03 ± 29.37 |
| NumPy | 2.561e+04 ± 639.30 | 2.7e+04 ± 627.80 |

TABLE V
BENCHMARK RESULTS FOR SERIAL RUNS USING PRE-CONVERTED TYPES, AS DESCRIBED ON SPECIALIZED CONSTRUCTOR (M2).

| Method | mean (ms) | stddev (ms) |
|---|---|---|
| ctypes | 1.369e+03 ± 43.84 | 1.658e+03 ± 89.95 |
| cffi (setuptools) | 633.1 ± 5.638 | 1.257e+03 ± 8.142 |
| cffi (Maturin) | 638.2 ± 4.451 | 1.265e+03 ± 5.393 |
| PyO3 | 634.7 ± 3.036 | 1.256e+03 ± 6.077 |
| NumPy | 262.5 ± 14.35 | 1.594e+03 ± 27.82 |

## B. Chunked Executions

The second benchmark compares chunked executions. They start at a chunk size of $2^{10}$ up to $2^{18}$, with intermediate samples in the exponential range, as shown in Fig. 2 and Fig. 3. Table VI and Table VII list all results respectively.

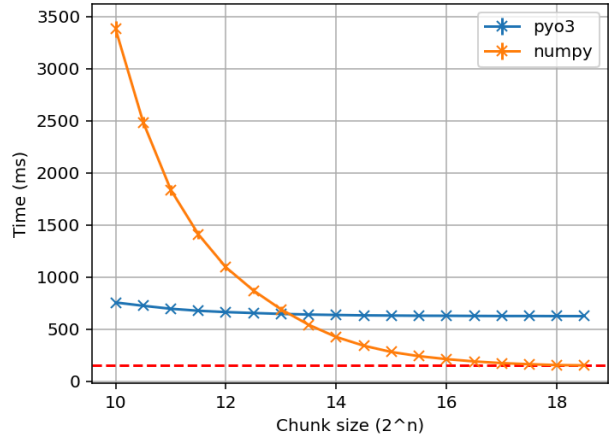

Fig. 2. **Chunked execution results for mean().** The sample size is divided in chunks and the execution times are summed for each implementation. The maximum error occurs at the least sample size of $2^{10}$ with 10.5 ms for PyO3 and 53.3 ms for NumPy (as shown in Table VI).

To measure the relative overhead against the NumPy calculations, the benchmarks subtract the minimum timing from the chunked results (shown as the dashed line on Fig. 2 and Fig. 3) and take them as a function of the number of calls – the sample size divided by the chunk size ($\frac{10^9}{2^n}$) – as shown in Fig. 4 and Fig. 5.

TABLE VI
BENCHMARK RESULTS FOR CHUNKED RUNS OF MEAN. THE VALUES ARE PLOTTED ON FIG. 2.

| Chunk size ($2^n$) | PyO3 | NumPy |
|---|---|---|
| 10.0 | 3.475e+03 ± 53.26 | 759.9 ± 10.56 |
| 10.5 | 2.51e+03 ± 27.40 | 722.4 ± 7.956 |
| 11.0 | 1.86e+03 ± 29.16 | 696.0 ± 6.109 |
| 11.5 | 1.421e+03 ± 14.28 | 677.6 ± 5.118 |
| 12.0 | 1.107e+03 ± 13.74 | 661.4 ± 4.354 |
| 12.5 | 870.1 ± 12.52 | 651.9 ± 3.613 |
| 13.0 | 682.9 ± 4.20 | 644.1 ± 2.497 |
| 13.5 | 546.7 ± 7.181 | 640.3 ± 4.441 |
| 14.0 | 425.7 ± 5.086 | 635.2 ± 4.231 |
| 14.5 | 343.8 ± 3.869 | 632.6 ± 2.779 |
| 15.0 | 286.1 ± 5.175 | 630.0 ± 3.308 |
| 15.5 | 244.0 ± 3.43 | 629.2 ± 3.210 |
| 16.0 | 212.9 ± 2.755 | 628.3 ± 2.759 |
| 16.5 | 190.4 ± 2.169 | 626.8 ± 3.805 |
| 17.0 | 173.8 ± 2.389 | 627.2 ± 3.346 |
| 17.5 | 165.5 ± 1.502 | 626.2 ± 3.150 |
| 18.0 | 154.9 ± 1.963 | 626.1 ± 1.835 |
| 18.5 | 153.2 ± 2.77 | 625.3 ± 2.010 |

TABLE VII
BENCHMARK RESULTS FOR CHUNKED RUNS OF STD. THE VALUES ARE PLOTTED ON FIG. 3.

| Chunk size ($2^n$) | PyO3 | NumPy |
|---|---|---|
| 10.0 | 1.348e+03 ± 11.59 | 9.382e+03 ± 64.00 |
| 10.5 | 1.316e+03 ± 5.457 | 6.808e+03 ± 61.71 |
| 11.0 | 1.296e+03 ± 5.033 | 5.040e+03 ± 46.21 |
| 11.5 | 1.285e+03 ± 8.125 | 3.825e+03 ± 53.61 |
| 12.0 | 1.278e+03 ± 3.923 | 2.928e+03 ± 22.35 |
| 12.5 | 1.272e+03 ± 5.900 | 2.258e+03 ± 12.12 |
| 13.0 | 1.264e+03 ± 8.286 | 1.791e+03 ± 14.13 |
| 13.5 | 1.258e+03 ± 4.963 | 1.466e+03 ± 8.84 |
| 14.0 | 1.253e+03 ± 3.192 | 1.233e+03 ± 6.972 |
| 14.5 | 1.253e+03 ± 3.767 | 1.071e+03 ± 9.129 |
| 15.0 | 1.251e+03 ± 2.522 | 958.8 ± 9.865 |
| 15.5 | 1.251e+03 ± 3.793 | 880.5 ± 9.754 |
| 16.0 | 1.250e+03 ± 2.485 | 828.2 ± 6.539 |
| 16.5 | 1.249e+03 ± 3.641 | 787.8 ± 3.932 |
| 17.0 | 1.249e+03 ± 3.290 | 765.5 ± 6.135 |
| 17.5 | 1.246e+03 ± 2.854 | 757.9 ± 4.599 |
| 18.0 | 1.248e+03 ± 2.348 | 772.5 ± 6.20 |
| 18.5 | 1.249e+03 ± 2.177 | 819.2 ± 7.03 |

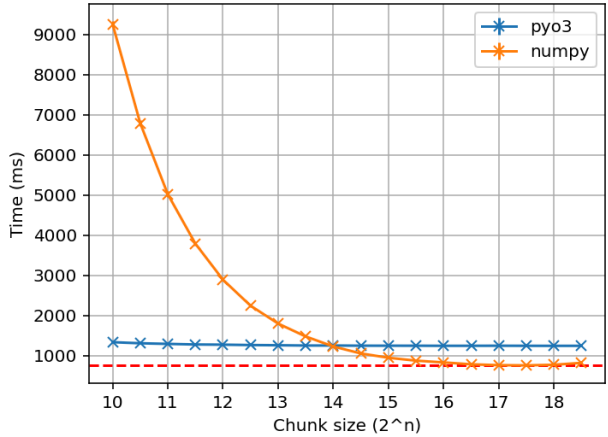

Fig. 3. **Chunked execution results for `std()`.** The sample size is divided in chunks and the execution times are summed for each implementation. The maximum error occurs at the least sample size of $2^{10}$ with 11.6 ms for PyO3 and 64.0 ms for NumPy (as shown in Table VII).

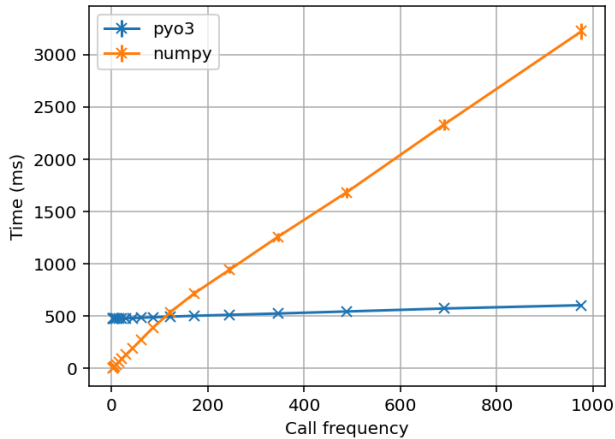

Fig. 4. **Function call overhead for `mean()`.** The least execution time for Numpy is subtracted from Fig. 2. Frequency measures # of function calls.

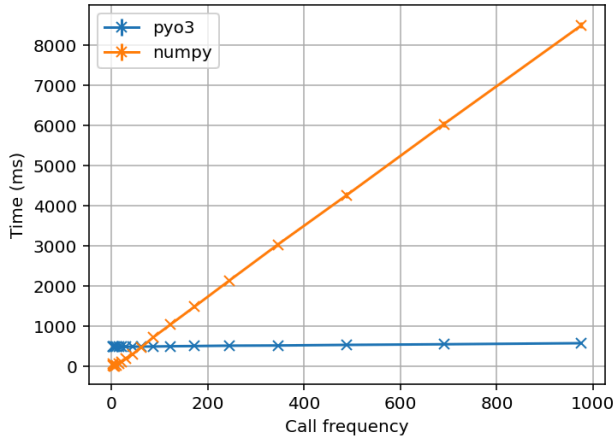

Fig. 5. **Function call overhead for `std()`.** The least execution time for Numpy is subtracted from Fig. 3. Frequency measures # of function calls.

TABLE VIII
LINEAR REGRESSION FOR FUNCTION CALL OVERHEADS.
PER-CALL OVERHEAD CORRESPONDS TO THE ANGULAR COEFFICIENT WHILE BASE SHOWS THE CONSTANT OVERHEAD AGAINST THE BEST NUMPY EXECUTION.

| Method | mean (ms) | | stddev (ms) | |
|---|---|---|---|---|
| | per-call | base | per-call | base |
| PyO3 | 0.1408 ± 0.001476 | 472.7 ± 0.3195 | 0.1017 ± 0.002395 | 1.095e+03 ± 0.5547 |
| NumPy | 3.562 ± 0.08204 | 14.36 ± 9.941 | 8.878 ± 0.06025 | 560.9 ± 8.442 |

Finally, the benchmark presents the linear regression results for the function call overheads, measured against the best case for `NumPy` (Fig. 4 and Fig. 5) in Table VIII.

## VII. ANALYSIS

As indicated in Section IV, `ctypes` has shown the most lacking alternative, requiring manual API redefinitions and expensive type constructions due to `libffi` outweighing any benefits as can be seen on Table IV. `cffi` allows easier performance gains and can also be integrated with Rust, making for a solid alternative for any libraries that already have shared library builds. `PyO3` provides the most flexibility when creating the Python bindings, which may be an advantage for library implementers, as they need workable APIs[5].

In Fig. 3, the `NumPy` implementation hits a minimum at a chunk size of $2^{16}$, which corresponds to 512KB of memory for 64 bit values, and also the machine L2 cache size[6], which indicates that their implementation makes full use of the machine resources. After rebounding from that minimum at the full sample size, the implementation outperforms `NumPy` by a small margin (seen on Table V) with no explicit SIMD instructions[7], which `NumPY` makes heavy use of.

Due to the difficulty of specifying flexible interfaces in standard C extension modules, `NumPy` defines its interfaces in pure Python, wrapping the appropriate low-level functions depending on e.g. value types or the data shape.

In both Fig. 4 and Fig. 5, the interpreter overhead caused by those interfaces contrasts with the higher base overhead, but lower per-call overhead in the custom implementations, as can be seen on Table VIII.

Table IX summarizes the analysis results regarding each of the objectives laid out on Section I.C.

## VIII. DISCUSSION

The previous sections evaluated the Rust-based `PyO3` toolchain, addressing its conciseness (RQ1), tooling (RQ2), and performance (RQ3). This section uses these results to answer the research questions described in Section I.

---

[5]The build systems are considered part of the tooling, thus it is irrelevant for the performance benchmark.
[6]Queried with `lscpu -C`.
[7]It is possible to instruct the Rust compiler to use specific optimizations, or inline assembly.

TABLE IX
Bindings Methods Analysis Summary.
All cffi accidental complexity can also be found on ctypes.

| Method | Conciseness RQ1 | Tooling RQ2 | Performance RQ3 |
|---|---|---|---|
| ctypes | Manual API (re-)declaration Manual type (re-)construction Requires libffi (dynamic symbol resolution) Inherits issues present on ctypes | Build system integration | Poor |
| cffi | Opaque struct exports Manual memory casts (unsafe) | Unclear documentation | Comparable |
| PyO3 | - | - | Comparable |

### A. RQ1 (Conciseness)

This research question was investigated through the implementations on Section IV. Comparing the code, PyO3's only issue in conciseness could come from the Rust language itself, which can be compared to C++ in many aspects. The library hides many of the lower-level details from the required C exports, while also preserving performance, which effectively makes it close a tooling gap for using Rust for this purpose.

### B. RQ2 (Tooling)

This research question was investigated through the implementations on Section IV. While this may be subjective, there are issues with both ctypes and cffi related to the legacy nature of the tools, i.e. because there is no standard tooling for C itself. However, both showed to be quite workable when the setup was correct, with most issues arising on the first-time setup. Gaps in tooling often appear under more specific circumstances, often in real-world scenarios.

### C. RQ3 (Performance)

This research question was investigated through the performance analysis Section VII. The experiments show that it is possible to outperform existing state-of-the-art libraries for specific use-cases, and ideally exposing the bindings without additional layers of abstraction, with the main takeaway that this can be done without Rust-specific speed-up techniques.

## IX. Threats to Validity

This section addresses the threats to validity related to the methodology described in Section III.

**Construct validity.** This paper does not address the challenge of switching from C to Rust, neither the actual implementation of lower-level optimizations. Moreover, the lack in variety of use cases can be an important limitation for properly comparing the toolchains.

**External validity.** This paper purposefully limited its scope to comparing only easily implementable mathematical functions. The results focus on the main aspects regarding FFI performance in favor of a complete optimization of specific algorithms.

## X. Conclusion

This research investigated the effectiveness of Rust-based tooling for developing Python libraries.

The results show that PyO3 offers ergonomic advantages in relation to C-specific tooling, such as cffi, with vast possibilities for optimizations. PyO3 offers higher-level tooling to bindings developers, and allows for Python-native interfaces with minimal per-call overhead. To optimize their bindings, developers should always focus on separating large type conversions and delineating memory boundaries.

For future work, other data structure conversions might be investigated, comparing memory layout differences. Moreover, it could also be useful to instrument the Python interpreter and understand the actual overheads in detail.